\documentclass[superscriptaddress,groupedaddress,nofootnoteinbib,12pt]{article}
\usepackage{graphicx}  
\usepackage{dcolumn}   
\usepackage{bm}        
\usepackage{jcappub}

\newcommand{\be}{\begin{equation}}
\newcommand{\ee}{\end{equation}}

\def\ba{\begin{eqnarray}}
\def\ea{\end{eqnarray}}

\def\nn{\nonumber}

\def\d{\mathrm{d}}
\def\mn{_{\mu \nu}}
\def\mupn{^\mu_{\, \nu}}
\def\({\left(}
\def\){\right)}

\def\o{\omega}
\def\Op{\Omega_P}
\def\l{\ell}

\def\hB{\hat{\Box^{\!\! \!\! \phantom L}}}
\def\astrosun{\odot}

\def\rs{r_\star}
\def\mpl{M_{\rm Pl}}
\def\p{\partial}

\begin{document}

\title{Vainshtein Mechanism in Binary Pulsars}

\author{Claudia de Rham$^{a}$, Andrew J. Tolley$^a$, Daniel H. Wesley}
\affiliation{$a$ Department of Physics, Case Western Reserve University, 10900 Euclid Ave, Cleveland, OH 44106, USA
\\
\\
{\bf E-mail:} claudia.derham@case.edu, andrew.j.tolley@case.edu}

\abstract{
We compute the scalar gravitational radiation from a binary pulsar system in the simplest model that exhibits the Vainshtein mechanism. The mechanism is successful in screening the effect from scalar fields conformally coupled to matter, although  gravitational radiation is less suppressed relative to its general relativity predictions than static fifth forces effects within the pulsar system. This is due to a combination of two effects: firstly the existence of monopole and dipole radiation; secondly the Vainshtein suppression comes from the hierarchy of scales between the inverse frequency scale and the Vainshtein radius, rather than the orbital radius of the pulsar system. Extensions of these results will have direct relevance to infrared modifications of gravity, such as massive gravity theories, which are known to exhibit a Vainshtein mechanism. Generalization to Galileon models with higher order interactions are likely to provide stronger constraints.
}

\maketitle

\section{Introduction}

The discovery of cosmic acceleration has spurred a search for
consistent modifications of gravity in the infrared, \cite{Dvali:2007kt}.  Theories in which the graviton acquires a mass, either softly as in the Dvali-Gabadadze-Porrati (DGP) model \cite{DGP} or cascading gravity \cite{deRham:2007rw}, or a hard mass as in the newly developed ghost-free models of massive gravity \cite{dRGT} or their bigravity \cite{Hassan:2011zd} extensions, are
a promising class of such modifications.  These models commonly
include light scalar degrees of freedom which arise from additional
graviton helicity states. At the linear level, the scalar modes do not decouple in the massless limit, which is known as the
van Dan, Veltman, Zakharov (vDVZ) discontinuity, \cite{van Dam:1970vg}.
Despite this, massive gravity models satisfy standard tests of
gravity since the new scalars become strongly coupled near dense
sources, which suppresses scalar gradients, and yields a force that
is much smaller than the Newtonian one. This strong coupling effect is known as the Vainshtein mechanism, \cite{Vainshtein}.
Far from compact sources,
the scalars are weakly coupled and massive gravity theories make
novel predictions that differ from Newtonian gravity in interesting
ways.  This suggests that IR modifications of gravity have extremely
small effects on any but cosmological scales.

Much of the phenomenology of the Vainshtein mechanism can be captured by considering the Galileon models \cite{Galileon}. Indeed the original Galileon model came from considering the decoupling limit of the DGP model \cite{lpr}, and the structure of this scalar was generalized in \cite{Galileon} to include all interactions consistent with the Galilean symmetry $\pi \rightarrow \pi + c + v_{\mu}x^{\mu}$, having a well-defined Cauchy problem. It was subsequently shown that the generic Galileon arises as the decoupling limit \cite{deRham:2010ik,deRham:2010gu} of the generic ghost-free massive gravity theory \cite{dRGT}. In an independent line of reasoning, the Galileon models have been viewed as scalar theories in their own right (independent of their graviton helicity-zero origin) and the most general covariant Galileon theories have been constructed \cite{CovariantGalileon,Deffayet:2009mn,deRham:2010eu,Goon:2011qf}. In the decoupling limit in which the helicity-zero graviton mode is viewed as weak, all of these theories take on a similar structure and at least in some cases, the physics of the Vainshtein mechanism is qualitatively similar.
For this reason, in this article we shall focus on the simplest cubic Galileon model, and will leave generalizations to subsequent work \cite{PulsarTheReturn}.
See Refs.~\cite{Wyman:2011mp,Koyama:2011xz,Chkareuli:2011te,Sjors:2011iv} for studies investigating the Vainshtein mechanism directly in massive gravity and Refs.~\cite{Babichev:2009jt} for Fierz-Pauli massive gravity. For other potentially related observations of the Vainshtein mechanism, see Ref.~\cite{Hui:2012jb,Iorio:2012pv,Babichev:2011iz}.

In the simplest cubic Galileon (DGP) model, the predicted anomalous acceleration $\Delta a_{\rm DGP}$ is
$\Delta a_{\rm DGP}/a_N  \sim
10^{-15}$, where $a_N$ is the Newtonian acceleration, for a
typical binary pulsar system of characteristic mass
$3 M_{\astrosun}$ and semi-major axis $a \sim 10^{-2}$ AU. We show that this ratio does not
set the anomalous contribution to radiated power, which is a few orders of magnitude larger in known binary pulsar systems and could potentially be even stronger in slow, high eccentricity systems.  The enhancement is due to the fact that
the relevant length scale is set by the onset of the ``far-field"
region at radius $\Omega_P^{-1}$ where $\Omega_P = 2\pi/T_P$ and $T_P$
the orbital period (often denoted $P_b$ in the literature.)  For $T_P =$ 8 hours, this is $\sim 10^3$ greater than the orbital radius. Furthermore, the system also radiates in the monopole and dipole channels since we are now radiating into a scalar.  Monopole and dipole radiation exists because the scalar effectively violates the equivalence principle. These effects could potentially get more pronounced in future observations.

Nevertheless, in the simplest model exhibiting the Vainshtein mechanism we consider in this paper, these enhancement are not sufficient to produce scalar gravitationally radiation effects which
are within present observational limits from pulsar timing, and the Vainshtein mechanism, albeit slightly more subtle in this time-dependent system, is still very much alive.
In a subsequent work we will show how higher order interactions in these different massive gravity/Galileon models are potentially much more strongly constrained by current observations \cite{PulsarTheReturn}.

The rest of this paper is organized as follows: In section \ref{sec:Formalism} we review the  cubic Galileon model and derive the general formalism to compute the radiated power in all generality in any multipole, taking care of the Vainshtein mechanism. The power is computed in two independent and equivalent ways using first an effective field theory approach and second a more conventional energy flux derivation. We then apply this formalism in the subsequent sections \ref{sec:Monopole}, \ref{sec:Dipole} and \ref{sec:quadrupole} to compute the respective monopole, dipole and quadrupole radiation in the cubic Galileon model. Finally we compare these results with observations in section \ref{sec:observations} and summarize our results.


\section{Pulsar Radiation in Galileon Model}

\label{sec:Formalism}
\subsection{Cubic Galileon}
Our starting point will be the so-called cubic Galileon model. This is the simplest model exhibiting the Vainshtein screening mechanism which was originally derived from the decoupling limit of the DGP model. In order to determine the emitted power it is sufficient to work in the so-called decoupling limit $\mpl \rightarrow \infty$, in which the self-interactions of the helicity-two graviton are neglected. For this reason it will not be of importance whether the scalar field is viewed to arise from a Galileon model in its own right \cite{Galileon}, DGP \cite{DGP}, massive gravity \cite{dRGT}, or some other infrared modification. Nevertheless just to be explicit on our conventions, if we were dealing with the Galileon model the form of the action would be
\be
S = \int \d^4 x \sqrt{-g}\left[  \frac{\mpl^2}{2} R  - \frac{3}{4} (\p \pi)^2 \left(1+ \frac{1}{3 \Lambda^3} (\Box \pi )\right) \right] ,
\ee
where $\Lambda$ is the strong coupling scale. In models of massive gravity, the scale $\Lambda$ is associated to the graviton mass $m$ by the relation $\Lambda^3=(m^2 \mpl)$.
It is usual (although not essential depending on the context) to assume that the scale $m$ is connected to the current Hubble scale $m\sim H_0\sim 1.54 \times 10^{-33}$eV, giving rise to a strong coupling scale $\Lambda^3 \sim  \mpl H_0^2\sim (1000 \, {\rm km})^{-3}$.

Writing the metric as $g\mn=\eta\mn +\mpl^{-1} h\mn$, we can work in the decoupling limit by sending $\mpl \to \infty$ keeping $\Lambda$ fixed.
Assuming a conformal coupling to matter for $\pi$, which is the case in all known theories of massive gravity (such as DGP, Cascading gravity or massive gravity), the action including matter is
\be
\label{decouplingaction}
S=\int \d ^4x \(-\frac{1}{4} h^{\mu\nu}({\mathcal{E}} h)\mn-\frac 3 4 (\partial \pi)^2\(1+\frac{1}{3 \Lambda^3}\Box
\pi\) +\frac{1}{2\mpl}
h^{\mu\nu}T\mn+\frac{1}{2\mpl}\,  \pi T\)\,,
\ee
where $({\mathcal E} h)_{\mu\nu}=-\frac{1}{2} \Box h_{\mu\nu}+\dots$, and $T$ is the
trace of the stress-energy tensor. As we can see, the helicity-2 and -0 mode decouple in that case.
The equations of motion for $\pi$ and $h\mn$ are then
\ba
\label{decoupledequations}
&({\mathcal E} h)_{\mu\nu}=\frac{1}{\mpl}T_{\mu\nu} \\
&\partial_{\mu}\(-\frac 3 2 \partial^{\mu}\pi
\(1+\frac{1}{3\Lambda^3} \Box \pi \) +\frac{1}{4\Lambda^3}
\partial^{\mu} \(\partial \pi \)^2\)=\frac{1}{2\mpl}T\, . \nn
\ea
Assuming a point source, $T_0{}\mupn = - M \delta^{(3)}(\vec{x})\delta^\mu_0 \delta^0_\nu$, the background solution for $\pi$ can be written simply if we introduce $E(r)$ and take $\vec\nabla\pi(r) =
\hat{r} E(r)$, \cite{Nicolis}.
Then the equations of motion are solved by
\ba
\label{e:SphericalEpm}
E_\pm(r) = \frac{\Lambda^3}{4r} \left[ \pm \sqrt{ 9r^4 + \frac{32 r_\star ^3 r}{\pi}} - 3r^2
\right]\,,
\ea
where the Vainshtein radius $r_\star$ associated with an object of mass $M$ is
\ba
r_\star=\(\frac{M}{16m^2 \mpl^2}\)^{1/3} = \frac{1}{\Lambda}\(\frac{M}{16\mpl}\)^{1/3}\,.
\ea
For $m\sim H_0$, an object with mass $1 M_\astrosun$ has  $r_\star  = 3\times 10^4$ pc.
We take the conventional branch solution $E_+$ which is trivial at infinity and bears no ghost-like instabilities.


\subsection{Perturbations}

The strong coupling effect is determined predominantly by the total
mass of the system. We split the stress energy as $T^{\mu\nu}=T_0^{\mu\nu}+\delta T^{\mu\nu}$ into a static `background' part $T_0^{\mu\nu}$ where we assume that the entire mass of the system is located at the center of mass ($\vec{x}=0$), and a perturbation $\delta T^{\mu\nu}$
which encodes the time dependent dynamics, which for slowly moving sources is
\ba
\label{e:stressenergy}
\delta T^\mu_\nu =
-\left[\sum_{i=1,2}M_i \delta^{(3)}(\vec{x}-\vec{x}_i(t))
-M \delta^{3}(\vec{x})\right]\delta^\mu_0\delta^0_\nu\,,
\ea
where $M_i$ is the mass of each companion and $M=M_1+M_2$.

Having decomposed the source in a background plus perturbations, we similarly split
$\pi=\pi_0+\sqrt{2/3}\, \phi$ and
express the quadratic Lagrangian for $\phi$,
\ba
\label{e:PhiAction}
\mathcal{L}_\phi &=& \frac 12
\(1 + \frac{2}{3\Lambda^3}\( E' + \frac{2E}{r} \) \)
\dot\phi^2-\frac 12 \( 1 + \frac{4}{3\Lambda^3}\frac{E}{r}\)
(\partial_r \phi)^2 \\
&-&
 \frac 12 \( 1 + \frac{2}{3\Lambda^3}\( E' + \frac{E}{r} \) \)
(\nabla_\Omega \phi)^2+\frac{\phi}{\sqrt{6}\mpl}  \delta T \,.\nn
\ea
This quadratic action will be sufficient to obtain our result for the power emission into scalar gravitational radiation. For pedagogical reasons we shall compute the power in two ways and demonstrate the consistency of the calculational results.

\subsection{Galileon Radiation}

\subsubsection{Method I -- Effective action}

In the effective action approach to the calculation of
gravitational radiation \cite{Goldberger:2007hy}, we first derive an
effective action for the dynamics of the matter distribution by
integrating out the graviton and scalar particles in the decoupling
action (\ref{decouplingaction}), giving the non-local matter
effective action expressed in terms of the Feynman propagator
\ba
S_{\rm{eff}}&=&\int \d^4 x {\mathcal L}_M
+\frac i{12\mpl^2}\int \d^4 x\d^4 x' \delta T(x)G_F(x,x')\delta
T(x')   \\ &+& \text{usual helicity-two contributions from GR}\,.\nn
\ea
Here we have used the fact that the field $\phi$ can be expressed in terms of the Feynman propagator
\ba
\phi(x)=\frac{i}{\sqrt{6} \mpl}\int \d^4x' G_F(x,x')\delta T(x')\,,
\ea
where we have defined the Feynman propagator via
\ba
\hB G_F(x,x')=i \delta^4(x-x')\,.
\ea
The modified d'Alembertian is the one appropriate to the quadratic action (\ref{e:PhiAction})
\ba
\hB \phi &=& -\(1 + \frac{2}{3\Lambda^3}\( E' + \frac{2E}{r} \) \)
\ddot \phi+\frac{1}{r^2} \frac{\partial}{\partial r} \( r^2 \( 1 + \frac{4}{3\Lambda^3}\frac{E}{r}\)
\partial_r \phi \) \\
&&+ \( 1 + \frac{2}{3\Lambda^3}\( E' + \frac{E}{r} \) \) \nabla_\Omega^2 \phi\,.\nn
\ea
As usual, the Feynman propagator can be expressed in terms of Wightman functions
\ba
G(x,x')=\theta(t-t')W^+(x,x')+\theta(t'-t)W^-(x,x')\,,
\ea
where
\ba
W^+(x,x')=\sum_{\l m}\int_0^{\infty} \hspace{-8pt}\d \o \, u_{\l m} (r,\Omega) u^\star_{\l m} (r',\Omega') e^{-i\o
(t-t')}
\,,
\ea
and the $u_{\l m}(r,\Omega)e^{-i\omega t}=u_{\l}(r)Y_{\l m}(\Omega)e^{-i\omega t}$ are a complete set of mode
functions in spherical harmonic space which satisfy the homogenous
equations of motion following from action~(\ref{e:PhiAction}).

The time-averaged power emission $P$ is
\ba
\label{e:Power}
P=-\left < \frac{\d \mathcal E}{\d t} \right >= \int_0^\infty \d  \o \, \o
f(\o)\,,
\ea
where $f(\omega)$ is determined from the imaginary part of the effective action
\ba
\frac{2\text{Im} S_{\rm {eff}}}{T_P}=\int_0^\infty \d  \o f(\o)\,,
\ea
and $S_{\rm eff}$ is calculated over one period.
We define the moments
\ba
\label{moments}\label{e:Mlmn}
M_{\l m n} =
\frac{1}{T_P} \int_0^{T_P}\hspace{-5pt} \d t \, \d^3 x\,
u_{\l m}(r,\Omega) e^{-i n \Omega_P t} \delta T \,.
\ea
Taking the Fourier transform
\ba
M_{\l m}=\sum_{n=-\infty}^\infty M_{lmn} e^{in \Op t}\,,
\ea
we have
\ba
f(\o)& = &\frac{1}{3\mpl^2 T_P} {\rm Re}\left[\sum_{\l m}
\int_0^{T_P} \hspace{-10pt}\d t \int_{-\infty}^{t}\hspace{-10pt} \d t'
e^{-i \o (t-t')} M_{\l m}(t) M^\star_{\l m}(t')\right]\\
 &=& \frac{\pi}{3\mpl^2}\sum_{n=0}^\infty \sum_{\ell, m} |M_{lmn}|^2 \, \delta(\o-n\Op)\,,\hspace{10pt}{\rm for }\ \o>0\,.\nn
\ea
From \eqref{e:Power}, the period-averaged power emission is then
\ba\label{e:Power4Multipoles}
P=\frac{\pi}{3\mpl^2}\sum_{n=0}^\infty
\sum_{\l m} \, n\Op  |M_{\l m n}|^2\,.
\ea
Inside the strong coupling region $r\ll r_\star$, the mode functions are approximately
\ba
u_\ell (r) = \bar u \(\frac{r}{r_\star}\)^{1/4} J_{\nu^\star} \(\frac{\sqrt 3}{2} \o
r\)\,,
\ea
with
\ba
\nu^\star=\left\{
\begin{array}{ccl}
(2\l+1)/4 & {\rm for} &  \ \ \l>0  \\
-1/4 & {\rm for} &\ \ \l=0
\end{array}
\right. \,,
\ea
and where we require the mode to be regular at $r=0$. This boundary condition is actually ambiguous for the monopole, and the monopole solution is determined by requiring that the first derivative vanishes $u_0'(r=0)=0$ so that this function is analytic in Cartesian coordinates.
The constant
$\bar u$ is fixed by imposing the correct normalization for the Feynman propagator
\ba
\hB G_F(x,x')=i \delta^4(x-x')\,.
\ea
Since $\hB W^+=0$, we have
\ba
\hB  G_F(x,x') &=&-\frac{2}{\sqrt{2\pi}}\(\frac{r_\star}{r}\)^{3/2}\Big[\delta(t-t')\partial_t W^+(t,t')
-\delta(t'-t)\partial_t W^-(t,t')\Big]\\
&=&-\frac{2}{\sqrt{2\pi}}\(\frac{r_\star}{r}\)^{3/2}\delta(t-t')\Big[\partial_t
W^+(t,t')-\partial_t W^-(t,t')\Big]\,.
\ea
We should therefore have
\ba
\sqrt{\frac{2}{\pi}}\(\frac{r_\star}{r}\)^{3/2} \left[\partial_t
W^+(t,t')-\partial_t W^-(t,t')\right]_{t=t'}=-i \delta^3(x-x')\,,
\ea
with
\ba
&&\sqrt{\frac{2}{\pi}}\(\frac{r_\star}{r}\)^{3/2}\left.\partial_t W^+(t,t')\right|_{t=t'}
=-i\sqrt{\frac{2}{\pi}}\(\frac{r_\star}{r}\)^{3/2}\sum_{\l,m}\int_0^{\infty}
\d \o \o u (r) u (r')
Y_{\l,m}(\theta,\phi) Y^\star_{\l,m}(\theta',\phi') \nn \\
&&=-i\bar u^2
\sqrt{\frac{2}{\pi}}\frac{r_\star}{r^{3/2}}
\sum_{\l,m}Y_{\l,m}(\theta,\phi)
Y^\star_{\l,m}(\theta',\phi')\(\int_0^{\infty} \d \o \o \(r r'\)^{1/4}
J_{\nu^\star} \(\frac{\sqrt 3}{2} \o r\)J_{\nu^\star} \(\frac{\sqrt 3}{2} \o
r'\)\)     \nn \\
&&=-i\bar u^2 \sqrt{\frac{2}{\pi}}\frac{r_\star
\(r r'\)^{1/4}}{r^{3/2}}\frac{4}{3}\sum_{\l,m}Y_{\l,m}(\theta,\phi)
Y^\star_{\l,m}(\theta',\phi')\(\int_0^{\infty} \d q q
J_{\nu^\star} \(q r\)J_{\nu^\star} \(q r'\)\)    \nn \\
&&=-i\bar u^2 r_\star \sqrt{\frac{2}{\pi}}
\frac{4}{3}\frac{\delta(r-r')}{r^2}\sum_{\l,m}Y_{\l,m}(\theta,\phi)
Y^\star_{\l,m}(\theta',\phi')   =-i\bar u^2 r_\star \sqrt{\frac{2}{\pi}}
\frac{4}{3}\delta^3(x-x')\,.
\ea
Consequently
\ba
\sqrt{\frac{2}{\pi}}\(\frac{r_\star}{r}\)^{3/2} \left[\partial_t
W^+(t,t')-\partial_t W^-(t,t')\right]_{t=t'}=
-2i\bar u^2 r_\star \sqrt{\frac{2}{\pi}}
\frac{4}{3}\delta^3(x-x')\,,
\ea
and the Greens function has the correct normalization for
\ba
\bar u^2=\frac 3 8\sqrt{\frac{\pi}{2}}\frac{1}{r_\star}.
\ea
Inside the strong coupling radius, the properly normalized mode functions are therefore
\ba
u_\l(r)=\(\frac{9\pi}{128}\)^{1/4}\frac{1}{\sqrt{r_\star}} \(\frac{r}{r_\star}\)^{1/4} J_{\nu^\star} \(\frac{\sqrt 3}{2} \o
r\)\,.
\ea
We can use the WKB approximation to extend this solution out to $r>r_*$,
\ba
\label{e:uWKB}
u_\l^{WKB}=\left\{\begin{array}{ccc}
\(\frac 3{8 \pi} \frac{1}{r \, r^3_\star \o^2}\)^{1/4} \cos \(\frac{\sqrt 3}{2} \o
r\) & \text{for} & \o^{-1}\ll r\ll r_\star\\
\frac{1}{\sqrt{\pi \o}} \frac{1}{r}\cos \( \o r\) & \text{for} & r\gg r_\star\\
\end{array}
\right.\,,
\ea
which corresponds to the correct field normalization well outside
the Vainshtein regime, where the field is living in flat
space-time. For $r\o \ll 1$, we are  well-inside the strong coupling region, and the mode functions can be expanded as
\ba
\hspace{-10pt}
u_\l (r)&\simeq&
\(\frac{9\pi}{128}\)^{1/4}\frac{1}{\sqrt{r_\star}}\(\frac{r}{r_\star}\)^{1/4}
(r\o)^{\nu^\star}\frac{3^{\nu^\star/2}}{4^{\nu^\star}} \left[ \frac{1}
{\Gamma(1+\nu^\star)}-\frac{3 }{16\Gamma(2+\nu^\star)}\, (r\o)^2 \right] \,.\hspace{10pt}
\ea
Using these in \eqref{e:Mlmn} we can then derive the power emitted from \eqref{e:Power4Multipoles}, which we apply in various cases after giving an
alternative and equivalent derivation of the radiated power.


\subsubsection{Method II -- Energy flux}

Our second technique involves computing the energy flux for the $\pi$ field directly.  In a diffeomorphism-invariant theory there is no local definition of the stress energy of the gravitational field.  One can define
a pseudo-tensor $t^{\mu\nu}$ such that
\ba
\partial_{\mu}\left (-g \(T^{\mu\nu}+t^{\mu\nu} \) \right
)=0\,,
\ea
where $T^{\mu\nu}$ is the stress energy of non-gravitational fields.
The conservation condition is satisfied by
$
t^{\mu\nu}=\tilde{t}^{\mu\nu}_{LL}+t^{\mu\nu}_{\pi},
$
where $\tilde{t}^{\mu\nu}_{LL}$ is the usual Landau-Lifshitz pseudotensor for the metric $\tilde{g}_{\mu\nu}$ expanded to second order, and $t^{\mu\nu}_{\pi}$ is the stress energy of the $\pi$ field
\ba
t_{\mu\nu}^{\pi} &=&   \frac{3}{2}\( \partial_{\mu} \pi
\partial_{\nu} \pi-\frac{1}{2} g_{\mu\nu}(\partial
\pi)^2\)+\frac{1}{2\Lambda^3}\partial_{\mu}\pi \partial_{\nu}\pi
\Box \pi \\ \nn
&-& \frac{1}{4\Lambda^3}\(
\partial_{\mu}(\partial \pi)^2 \partial_{\nu}\pi+
\partial_{\nu}(\partial \pi)^2
\partial_{\mu}\pi-g_{\mu\nu}
\partial_{\alpha}(\partial \pi)^2 \partial^{\alpha}\pi\).
\ea
Since the stress-energy $t\mn$ splits into two decoupled terms, and the first contribution gives the standard Einstein gravity result, we need calculate only the additional flux from the $\pi$ field.
Expanding around the background solution $\pi_0$, the radial energy flux gives
\be
t^\pi_{0r} = \(1 + \frac{4}{3\Lambda^3} \frac{E}{r}  \)\partial_t \phi \partial_r
\phi\,,
\ee
up to irrelevant total derivative terms.   The stress-energy is conserved, so we can find the total power emission by integrating
the flux over a sphere of any convenient radius.
The solution for an outgoing wave in the region $r \gg \Omega_P^{-1}$ is
\be
\phi=\sum_{\l mn}\phi_{\l m n}(r)Y_{\l m}e^{in\Omega_P t} \, ,
\ee
 where
\be
\phi_{\l m n}=\frac{\sqrt{c_s}\ a_{\l m n}}{r\sqrt{1 + \frac{4}{3\Lambda^3} \frac{E}{r} }}e^{-in \Omega_P \int c_s^{-1} dr},
\ee
with the wave propagation speed $c_s = 2/\sqrt{3}$ in the strong coupling region.
From this we infer the period-averaged power emission of
\be
\label{power2}
P=\hspace{-5pt}\sum_{n=-\infty}^\infty \hspace{-2pt}\sum_{\l m} n^2 \Omega_P^2 |a_{\l m n}|^2=2\sum_{n=0}^{\infty} \sum_{\l m }n^2 \Omega_P^2 |a_{\l m
n}|^2.\hspace{-6pt}
\ee
From the equation of motion for $\phi$ it is straightforward to derive the Wronskian identity
\be
\nn
\lim_{r \gg \Omega_P^{-1}} \left[r^2 \(1+\frac{4E}{3\Lambda^3 r}\)\(u_{l,r}\phi_{\l mn} -u_l \phi_{\l mn,r}\)\right]=\frac{M_{\l mn}}{\sqrt{6}\mpl},
\ee
from which we infer that
\be
|a_{\l mn}|^2=\frac{\pi}{6|n|\Omega_P}\frac{1}{\mpl^2}|M_{\l
mn}|^2\,,
\ee
 for $n\neq0$. On substitution into Eq.~(\ref{power2}) this reproduces exactly the result of Eq.~\eqref{e:Power4Multipoles}. Having derived the general formula for the power emitted by the Galileon in two independent ways, we now turn to computing the explicit low multipole radiation.


\section{Power Emission in Monopole}
\label{sec:Monopole}
To determine the power emission, we compute the multipole coefficients $M_{\ell m n}$ and use Eq.~(\ref{e:Power4Multipoles}). The Keplerian orbit of the two masses in the binary system is
\ba
\label{r(t)}
r_{1,2}(t)=\frac{\bar r (1-e^2)}{1+e \cos \Omega_P t}
\frac{M_{2,1}}{M}\,,
\ea
where
$e$ is the orbital eccentricity, and $\bar r$ the semi-major axis.
In particular, near the pulsar, the zero mode function behaves as
\ba
\label{u_0}
u_0 (r)&\simeq&
\frac{\beta}{(\o r^3_\star)^{1/4}}\(1-\frac{(\o r)^2}{4}\) \, ,
\ea
where
\ba
\beta=\(\frac{9\pi}{128}\)^{1/4}\frac{3^{-1/8}\sqrt{2}}{\Gamma (3/4)}\simeq 0.69\,.
\ea

\subsection{Relativistic Corrections}

It transpires that as in GR, the leading order contribution to the monopole and dipole formula will vanish as a consequence of energy and momentum conservation despite the fact that the equation for the scalar fluctuations is strongly Lorentz violating in the Vainshtein regime. For this reason we need the mode functions to subleading order for the monopole and dipole to obtain a non-zero estimate for the power emitted in these modes. Given this fact we should work with the first relativistic corrections that arise to the trace of the stress energy. To first order in relativistic corrections we have
\be
\delta T =-\left[
\sum_{i=1,2}M_i \left( 1-\frac{1}{2}v_i^2- \Phi_i  + \dots \right) \delta^{(3)}(\vec{x}-\vec{x}_i(t))-M \delta^{(3)}(\vec x)\right]\,,
\ee
where $\Phi_{i}$ are the Newtonian potentials evaluated at the location of each source. However as in GR, to second order in velocity, the conserved non-relativistic energy $E_{\rm NR}$ is simply given as  $E_{\rm NR}=\sum_i ( \frac 12 M_i v_i^2 -M_i \Phi_i )$, so that we can express $\sum_i M_i(\frac 12 v_i^2+\Phi_i)=\sum_i M_i v_i^2-E_{\rm NR}$ to that order in velocity.
The velocity includes a radial and angular contribution,  $v^2_i=\dot r_i^2(t)+(\o r_i(t))^2$ but using \eqref{r(t)}, we can check that the radial velocity is suppressed by two powers of the eccentricity compared to angular velocity $(\o r_i(t))^2$. Since the eccentricity is of order $e\approx 0.1 - 0.6$ in realistic binary pulsar models, we can ignore their contribution, and use the approximation $\sum_i M_i(\frac 12 v_i^2+\Phi_i)\simeq \sum_i M_i (\o r_i(t))^2-E_{\rm NR}$.
To second order in the velocity, the trace of the stress-energy tensor thus include the relativistic corrections,
\ba
\delta T =-\left[\sum_{i=1,2}M_i \left( 1-(\o r_i(t))^2 \right) \delta^{(3)}(\vec{x}-\vec{x}_i(t))-(M-E_{\rm NR}) \delta^{(3)}(\vec x)\right]\,.
\ea
Coupling the mode functions to this trace, therefore leads to the following contribution up to second order in $\o r$,
\ba
\label{CouplingMonopole}
\int \d^3 x u_0(r) \delta T &=& - \frac{\beta}{(\o \rs^3)^{1/4}}\left[\sum_{i=1,2}M_i \(1 -\frac 54 (\o r_i(t))^2\) -(M-E_{\rm NR})\right] \\
&=&\frac{5\beta}{4(\o \rs^3)^{1/4}} \sum_{i=1,2}M_i (\o r_i(t))^2 + \text{non-radiating contribution}\,,
\ea
where we ignore the piece going as $E_{\rm NR}$ as it is conserved and thus cannot lead to any radiative contribution (time-independent contributions scale as $\delta_{n0}$ in the monopole moment $M_{00n}$ and thus do not affect the power which goes as $\sum_{n\ge 0} n |M_{00n}|^2$. This is to be expected since static sources cannot radiate).

\subsection{Power}

Defining the effective monopole mass
\ba
\mathcal{M}_M=\frac{M_1M_2^2+M_2M_1^2}{M^2}\,,
\ea
and using the expressions \eqref{CouplingMonopole} and \eqref{r(t)} in the relation \eqref{e:Mlmn} for the monopole moment, with $\ell=m=0$ gives
\ba
M_{00n}&=&\frac{5\beta}{4(\o r^3_\star)^{1/4}}\frac{(\o\bar r)^2}{T_P}\mathcal{M}_M\,
(1-e^2)^2 \int_0^{T_P}\d t \frac{e^{-in\Op t}}{(1+e \cos (\Op t))^2} \,,
\ea

The fact that it is the subleading term in the mode expansion that gives the main effect is a statement that the leading order contribution to the monopole radiation vanishes due to conservation of energy/mass, which is also why there is no monopole radiation in GR. The difference here is that because the source is $\delta T$ and not the exactly conserved energy, there are relativistic corrections which lead to monopole radiation.
Defining
\ba
\label{IMn}
I_n^M(e) & = & \frac{n^{9/4}}{2 \pi} (1-e^2)^2 \int_0^{2 \pi} \frac{e^{- i n x}}{(1+e \cos x)^2} \d x\\
&=& n^{9/4}\sqrt{1-e^2} e^{-n} (\sqrt{1-e^2}-1)^n (1+n \sqrt{1-e^2})\, ,
\ea
the expression for the total power emitted via the monopole is hence
\ba
P_{\rm monopole}&=&\frac{\pi}{3\mpl^2}\sum_{n\ge 0} n\Op  |M_{00n}|^2\\
&=& \(\frac{25 \pi}{3 }\frac{\beta^2}{16}\) \frac{(\Op \bar r)^4}{(\Op r_\star)^{3/2}}
 \frac{\mathcal{M}_M ^2}{\mpl^2} \Op^2 \sum_{n > 0}|I_n^M(e)|^2\,,
 \label{e:TheAnswer}
\ea
This integral $I_n^M(e)$ is highly dominated by the low-$n$ harmonics (till $n=10$ as can be seen from Fig.~\ref{fig:Integrals}). In principle one could include the sum of all the $n$'s till a maximal $n_{\max}= (\bar r \o)^{-1}$ at which point the approximation used in deriving the mode functions breaks down. However the contribution these extra harmonics is completely negligible.

Performing this sum over $|I_n^M(e)|^2$ numerically for a typical eccentricity of 0.617 corresponding to that of the Hulse-Taylor pulsar  \cite{Hulse:1974eb,Taylor:1989sw,Weisberg:2004hi}, we obtain,
\ba
P_{\rm monopole} &\simeq& 4 \frac{(\Op \bar r)^4}{(\Op r_\star)^{3/2}}
\frac{\mathcal{M}_M^2}{\mpl^2}\Op^2\,.
\ea

\subsection{Comparison with GR}
The formula (\ref{e:TheAnswer}) is the analogue of the Peters-Mathews formula
\cite{Peters:1963ux} for quadrupole gravitational radiation,
\be
P_{\rm Peters-Mathews} = \frac{32}{5} \frac{G^5}{c^5} \frac{M_1^2 M_2^2 (M_1+M_2))}{\bar{r}^5} \frac{\left(1+\frac{73}{24} e^2+\frac{37}{96} e^4\right)}{(1-e^2)^{7/2}} \, ,
\ee
which using Kepler's third law $\Omega_P^2 = G(M_1+M_2)/\bar{r}^3$, $c=1$, $\mpl^2=1/8 \pi G$ can be re-expressed into current notation as
\be
\label{Peters-Mathews}
P_{\rm Peters-Mathews} = \frac{4}{5 \pi \mpl^2}\frac{\left(1+\frac{73}{24} e^2+\frac{37}{96} e^4\right)}{(1-e^2)^{7/2}} (\Omega_P \bar{r})^4 \frac{M_1^2M_2^2}{ (M_1+M_2)^2} \Omega_P^2 \, .
\ee
Comparing our two results we learn one important thing: The suppression factor that arises due to the Vainshtein effect is given by
\be
\text{Radiative Vainshtein suppression } = \frac{1}{(\Omega_P r_\star )^{3/2}} \, .
\ee
This should be directly compared with the typical suppression from fifth forces
\be
\text{Static (Fifth Force) Vainshtein suppression} = \left(\frac{\bar r}{r_{\star}} \right)^{3/2} \, .
\ee
In other words there is an enhancement of $v^{-3/2}\sim(\Op \bar r)^{-3/2}$
which indicates that, in this case, the  Vainshtein effect, whilst still fully active, is nevertheless slightly less powerful in this fully time-dependent situation. At the moment it is yet
unclear whether this result is generic in any time-dependent system or
whether there could be other configurations where the additional time
dependence provides an additional Vainshtein screening. This calls  for a
more general analysis, which lies beyond the scope of this work.

\section{Dipole Radiation}
\label{sec:Dipole}

We now turn to the dipole radiation for which the mode functions take the form in the small $(\o r)$ regime
\ba
u_1(r)\approx \frac{(\o r)}{(\o \rs^3)^{1/4}}\(\gamma_0-\gamma_1 (\o r)^2\)\,,
\ea
with
\ba
\gamma_0=\(\frac{9 \pi}{128}\)^{1/4}\frac{3^{3/8}}{4^{3/4}}\frac{1}{\Gamma(7/4)}\sim 0.4 \hspace{20pt}{\rm and}\hspace{20pt}
\gamma_1=\gamma_0\frac{3}{16}\frac{\Gamma(7/4)}{\Gamma(11/4)}\sim 0.04\,.
\ea

When computing the relativistic corrections to the trace of the energy momentum tensor, the situation is slightly different for the dipole than the monopole, and cannot easily be derived. Nevertheless, since the order of magnitude of the dipole is well below that of the monopole and the quadrupole, as it shall transpire, we can simply give an order of magnitude estimation and use,
\ba
\label{CouplingDipole}
\int \d^3 x u_1(r) \delta T =  -\frac{1}{(\o \rs^3)^{1/4}}\sum_{i=1,2}M_i \(c_0 (\o r_i(t)) -c_1 (\o r_i(t))^3\) \,,
\ea
where $c_0$ and $c_1$ are two constants with $c_1\sim \gamma_0+\gamma_1\sim 0.4$.

\subsection{Leading Contribution}

The  dipole moment $M_{1mn}^{(0)}$ corresponding to the leading contribution in $c_0 (\o r_i(t))$ is,
\ba
M_{1mn}^{(0)}=-c_0\frac{2 M_1 M_2}{M}\frac{(\o \bar r)}{(\o r^3_\star)^{1/4}}\frac{(1-e^2)}{T_P}\,
\int_0^{T_P}\hspace{-10pt}\d t \frac{e^{-in\Op t}}
{(1+e \cos (\Op t))}\sum_{i=1,2}Y_{1,m}(\theta_i,\phi_i(t)) \,,
\ea
where the angular position of the first object is taken to be localized at $\theta_1=\pi/2, \phi_1(t)=\Op t$, while the second object is diametrically opposed $\theta_2=\pi/2, \phi_2(t)=\Op t+\pi$. Using the fact that $Y_{1, 0}(\theta=\pi/2, \phi)=0$ for any angle $\phi$ and $Y_{1, \pm 1}(\pi/2,\phi)=-Y_{1, \pm 1}(\theta, \phi+\pi)$, we therefore see, as in the case of the monopole that the leading contribution cancels, which this time is a consequence of angular momentum conservation.

The leading order contribution to the dipole moment $M_{1mn}^{(0)}$ is the same as the dipole moment in GR, and this contribution vanishes for the same reason as in GR, as angular momentum is conserved. We have already seen this with the monopole, where the leading contribution vanishes because of conservation of energy (mass in this non-relativistic limit). The difference between the monopole and the dipole is that in the present case the subleading contribution to the monopole is parametrically larger than anticipated, whereas for the dipole we shall see that it is much smaller.

\subsection{Subleading Corrections}

The next order moment $M_{1mn}^{(1)}$  in the expansion in the expansion in $(\o r)$ gives a non-trivial but small contribution
\ba
M_{1mn}^{(1)}&=&c_1\frac{(\o \bar r)^3}{(\o r^3_\star)^{1/4}}\frac{(1-e^2)^3}{T_P}\,
\int_0^{T_P}\hspace{-10pt}\d t \frac{e^{-in\Op t}}
{(1+e \cos (\Op t))^3} \times \nn \\
&& \times \(\frac{M_1 M_2^3}{M^3}Y_{1m}(\frac{\pi}{2},\Op t)+\frac{M_2^3 M_2}{M^3}Y_{1m}(\frac{\pi}{2},\Op t+\pi)\) \,,
\ea
which is only relevant for for $m=\pm 1$ and for systems with significantly different masses which is not the typical case of binary pulsar systems.
Setting
\ba
\Delta M= \frac{M_1 M_2 (M_1^2-M_2^2)}{M^3}\,,
\ea
we get
\ba
M_{\ell=1, m=1 ,  n}^{(1)}(e)&=& -\frac{c_1}{2\sqrt{n}}\sqrt{\frac{3}{2\pi}}\frac{(\Op \bar r)^3}{(\Op r^3_\star)^{1/4}}\Delta M I^D_{n}(e)\,.
\ea
The $m=-1$ channel is suppressed compared to the $m=1$ as it picks up the opposite rotation direction. The main reason is that the $m=1$ contribution goes $e^{i m \phi}$ and thus cancels the oscillating behaviour of $e^{-i n \phi}$ for $n=1$, whilst the $m=-1$ never cancels the oscillation for positive $n$. An explicit computation, shows that $P_{\ell=1, m=-1}\lesssim10^{-2}P_{\ell=1, m=1}$, and thus only focus on the $m=2$ contribution in what follows.
The power emitted through this channel is thus
\ba
\label{e:PowerDipole}
P_{\rm dipole}=  \frac{c_1^2}{8 } \frac{(\Op \bar r)^6}{(\Op \rs)^{3/2}}\frac{\Delta M^2}{\mpl^2}  \Op^2 \sum_{n=0}^\infty \left|I^D_n(e)\right|^2\,,
\ea
with
\ba
\label{Idn}
I^D_n(e) & =& (1-e^2)^3\frac{n^{13/4}}{2\pi}\int_0^{2\pi} \frac{e^{-i(n-1)x}}{(1+e \cos x)^3}\d x\\
&=&\frac 12 n^{13/4}\sqrt{1-e^2}(\sqrt{1-e^2}-1)^{n-1}\\
&& \times (3-3\sqrt{1-e^2}-n^2 (e^2-1)+n(-2+2e^2+3\sqrt{1-e^2}))\,.\nn
\ea
Performing this sum over $I_n(e)$ numerically for a typical eccentricity of 0.617 corresponding to that of the Hulse-Taylor pulsar \cite{Hulse:1974eb,Taylor:1989sw,Weisberg:2004hi}, we obtain,
\ba
\label{sum_dipole_HT}
\sum_{n=0}^\infty \left|I^D_n(0.617)\right|^2 \approx 10^3\,,
\ea
which gives an overall dipole power emitted of
\ba
P_{\rm dipole}\approx 40 \frac{(\Op \bar r)^6}{(\Op \rs)^{3/2}}\frac{\Delta M^2}{\mpl^2}  \Op^2\,,
\ea
which, for the Hulse-Taylor pulsar is roughly 9 orders of magnitude below the monopole radiation and is thus utterly negligible. Notice that the situation could potentially change ever so slightly for systems with a bigger mass difference between the companions.

\section{Quadrupole Radiation}
\label{sec:quadrupole}

\begin{figure}
  \begin{center}
    \includegraphics[width=2in]{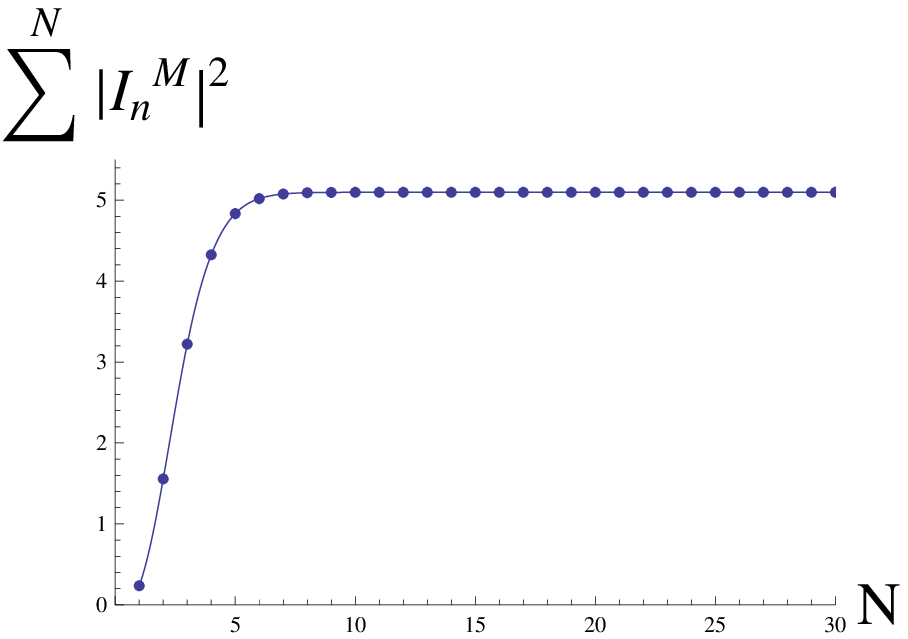} 
    \includegraphics[width=2in]{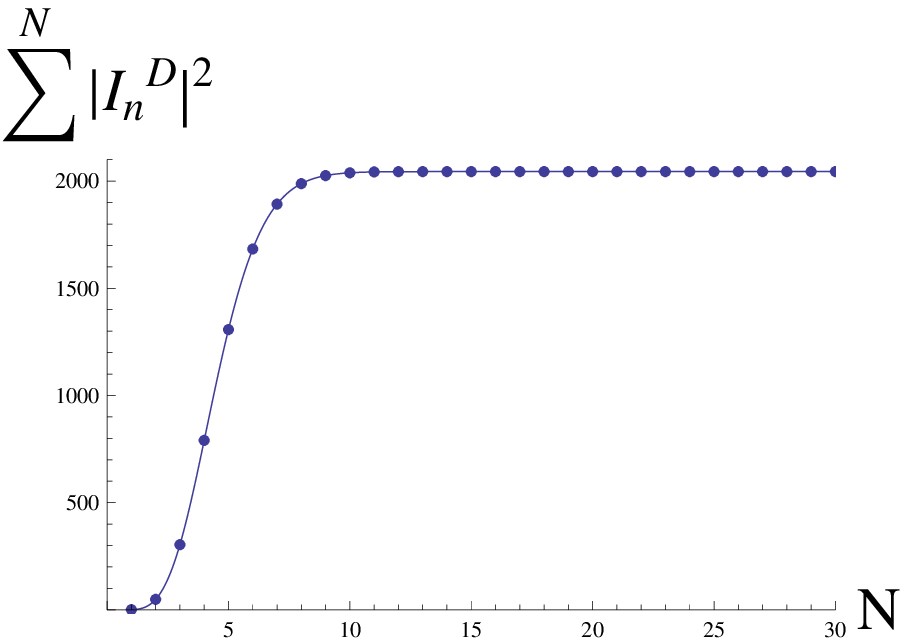}
    \includegraphics[width=2in]{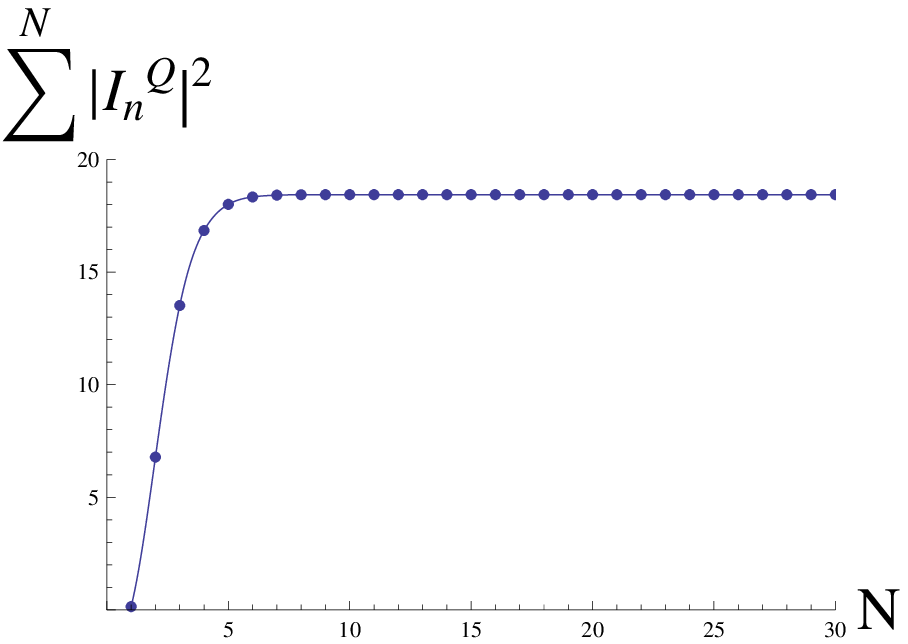}
  \end{center}
  \caption{Numerical evaluation of the the sums $\sum_{N=0}^\infty |I^M_n(e)|^2$, $\sum_{N=0}^\infty |I^D_n(e)|^2$ and $\sum_{n=0}^N |I^Q_n(e)|^2$ with $I^{M, D, Q}_n$ expressed in \eqref{IMn}, \eqref{Idn} and \eqref{IQn} respectively for the
  eccentricity of the Hulse-Taylor pulsar 1913+16 $e=0.617$,  \cite{Hulse:1974eb,Taylor:1989sw,Weisberg:2004hi}. The sums show a rapid convergence as $N\to \infty$. }
  \label{fig:Integrals}
\end{figure}

We finish by estimating the quadrupole emitted by the Galileon. In that case the mode function is of the form
\ba
u_2(r)\approx \lambda \frac{(\o r)^{3/2}}{(\o \rs^3)^{1/4}}\(1-\frac 1{12}(\o r)^2\)\,,
\ea
where
\be
\lambda = \frac{3^{9/8} \pi^{1/4}}{2^{17/4} \Gamma(9/4)}.
\ee
In this case the leading contribution to the mode functions gives the dominant contribution to the power and we can thus ignore the subleading and relativistic corrections. The moment vanishes for $m=\pm 1$.  Here again, out of the three other moments, $m=-2,0,+2$, the leading contribution arises from the $m=2$ one as it goes as $e^{i m \phi}$ and thus cancels the oscillating behaviour of $e^{-i n \phi}$ for $n=2$. For $m=0$ and $-2$ on the other hand this cancelation can never occur for $n>0$ and thus the contributions from these integrals is always suppressed compared to the $m=2$. An explicit computation, shows that $P_{\ell=2, m=0}\lesssim10^{-2}P_{\ell=2, m=2}$ and $P_{\ell=2, m=-2}\lesssim 10^{-3}P_{\ell=2, m=2}$, and thus only focus on the $m=2$ contribution in what follows. In that case
\ba
M_{\ell = 2, m=2 ,  n}(e)&=&-\frac{1}{4}\sqrt{\frac{15}{2\pi}} \frac{\lambda}{\sqrt{n}}\frac{(\Op \bar r)^3}{(\Op r^3_\star)^{1/4}} M_Q I^Q_{n}(e)\,,
\ea
with
\ba
M_Q=\frac{M_1 M_2\(\sqrt{M_1}+\sqrt{M_2}\)}{M^{3/2}}\,,
\ea
and
\ba
\label{IQn}
I^Q_{n}(e)=(1-e^2)^{3/2}\frac{n^{7/4}}{2\pi}\int_0^{2\pi} \frac{e^{-i(n-2)x}}{(1+e \cos x)^{3/2}}\d x\,.
\ea
The power emitted through this channel is thus
\ba
\label{e:PowerQuadrupole}
P_{\rm quadrupole}=  \frac{5\lambda^2}{32} \frac{(\Op \bar r)^3}{(\Op \rs)^{3/2}}\frac{M_Q^2}{\mpl^2}  \Op^2 \sum_{n=0}^\infty \left|I^Q_n(e)\right|^2\,.
\ea
Once again, the sum is dominated by the low-frequency harmonics $n\lesssim 15$ but the sum can be performed numerically for up arbitrarily large values, (see Fig.~\ref{fig:Integrals}). Fortunately it converges rapidly, and for the Hulse-Taylor pulsar  \cite{Hulse:1974eb,Taylor:1989sw,Weisberg:2004hi}, we obtain
\ba
\sum_{n=0}^\infty \left|I^Q_n(0.617)\right|^2 \approx 18\,,
\ea
which leads to a power emitted roughly 100 times larger than the monopole.

Comparing the quadrupole power emitted via the Galileon field with that emitted via the standard GR helicity-2 mode as given by the Peters-Mathews formula in Eq.~\eqref{Peters-Mathews}, we find a suppression factor given by
\be
\frac{P_{\rm quadrupole}^{\rm  Galileon}}{P_{\rm quadrupole}^{\rm  GR}}= q\  (\Omega_P r_\star )^{-3/2} (\Omega_P \bar r )^{-1} \, ,
\ee
up to a numerical factor $q$  which depends on the eccentricity of the system  and on the mass difference between the two objects (for the Hulse-Taylor pulsar, that factor is $q \simeq 0.08$, \cite{Hulse:1974eb,Taylor:1989sw,Weisberg:2004hi}). As one can see, this power is suppressed by one less power of the velocity compared to the GR result and could thus in principle have been important if it had not simultaneously been Vainshtein suppressed by  a factor of $(\Omega_P r_\star )^{-3/2}$. In this case the enhancement compared to the static spherically symmetric Vainshtein effect goes as $v^{-5/2}\sim(\Op \bar r)^{-5/2}$, so once again the Vainshtein mechanism is slightly less powerful than in the fully static configuration. Whether or not this is a generic statement should be explored more thoroughly.

\section{Discussion}
\label{sec:observations}

Pulsars in DNS
binaries are  used to measure gravitational-radiation
effects.  Timing measurements of these systems are  free from
contamination due to tidal effects or accretion from a stellar
companion. The archetypal such system is the Taylor-Hulse pulsar
1913+16 discovered in 1974,  \cite{Hulse:1974eb,Taylor:1989sw,Weisberg:2004hi}.  We have summarized the orbital parameters of four known DNS pulsars (A to D) and one pulsar-white dwarf binary (E)
in Table 1, \cite{Lorimer:2005bw},   all of which have measured orbital
period derivatives $\dot T_P$, which agree with GR.

The orbital period derivatives $\dot T_P$ are given in terms of the non-relativistic energy $E_{\rm NR}$ and the power emitted by the relation
\ba
\label{eq:TPdot}
\dot T_P =\frac{3}{2} T_P \frac{1}{E_{\rm NR} }\frac{\d E_{\rm NR}}{\d t}\,,
\ea
where the system's non-relativistic energy is
\ba
E_{\rm NR}=\frac{1}{2 (8\pi)^{2/3}}\frac{M_1 M_2}{\mpl} \(\frac{\Op^2}{M \mpl}\)^{1/3}\,.
\ea
The monopole and quadrupole Galileon radiation are close, but in the explicit examples presented here, the Galileon quadrupole always give the largest contribution. In summary, in the simplest model, the Galileon radiation is at least 7 orders of magnitude below that of GR (7 orders of magnitude for pulsar C and E and 8 for the other DNS pulsars),  which 6 orders of magnitude below the current precision, \cite{Footnote}, when considering a parameter $m=1.54 \times 10^{-33}$eV, or equivalently for a strong coupling scale $\Lambda\sim 10^{-13}{\rm eV}\sim (1000 {\rm km})^{-1}$. The best precision is still the Hulse-Taylor pulsar and pulsars C and E. Even though the double pulsar D has a good precision $\sigma$, its low eccentricity make it not the best candidate to probe the Vainshtein mechanism.

As can be seen from the expression \eqref{eq:TPdot} for the orbital period derivative, using the power emitted in the monopole, \eqref{e:TheAnswer}, the dipole \eqref{e:PowerDipole} or the quadrupole, \eqref{e:PowerQuadrupole}, $\dot T_P$ scales directly as $\rs^{-3/2}\sim m \sim \Lambda^{3/2}$.
So in order for the scalar field $\pi$ to have an effect at all on current or upcoming binary pulsar timing observations, the parameter $m$ should be enhanced by roughly 6 orders of magnitudes. Binary Pulsar timing thus put a rough bound of $m < 10^{-27}$eV or $\Lambda < 10^{-9}$eV. Compared to solar system tests \cite{solarsystem}, these bounds are not competitive for the cubic Galileon interactions, but could be enhanced when considering higher interactions, \cite{PulsarTheReturn}.

Present constraints are limited by the sample of DNS pulsar systems.
Ideal DNS binary pulsars have long periods, high eccentricities, and
are located nearby so that the kinematic corrections are reduced.
Over time, measurements of the orbital period derivative become more precise as more data is gathered.
The ultimate precision is limited by the uncertainty in the relative
acceleration between the sun and the pulsar system, which must
be included to obtain the intrinsic orbital period derivative from
the apparent one.

We should note that whilst previous authors have considered constraints on the mass of the graviton in binary pulsars, in particular see \cite{Finn:2001qi}, these authors do not take account of the Vainshtein mechanism, working only in the linearized Fierz-Pauli theory, and they further utilize an incorrect expression for the stress energy of gravitational radiation which accounts only for the helicity two component. As such the constraints obtained there are not appropriate for consistent Lorentz invariant theories of massive gravity as considered in \cite{dRGT}. They may however be relevant to theories in which there is no propagating helicity zero mode.

In this paper, we have shown the successful implementation of the Vainshtein mechanism in a fully time-dependent setup. This explains how a conformally coupled scalar field, or the helicity-0 mode of a massive graviton can evade typical tests of GR and particularly the well constrained orbital period decay in binary pulsar systems. Nevertheless, despite the presence of an active Vainshtein mechanism, we show that the suppression is less than naively anticipated from purely static systems for several reasons:
\begin{itemize}
\item The suppression factor that arises due to Vainshtein effect is going like $(\Op \rs)^{-3/2}$ for the monopole srather than $(\bar r/\rs)^{3/2}$ as is the case in static spherically symmetric configurations (for the quadrupole, the suppression factor even acquires an additional velocity suppression).
\item Furthermore, even though the leading contributions from the monopole and dipole radiation vanish from energy and angular momentum conservation, the sub-leading (relativistic) contributions are non-negligible, and in the case of the monopole can be comparable to the Galileon quadrupole radiation.
\item Finally, we have focused here on the simplest realization of the Vainshtein mechanism, namely within the context of the cubic Galileon model. However higher order Galileon interactions could potentially lead to an additional enhancement of the Galileon radiation and ought to be studied in their own right, \cite{PulsarTheReturn}.
\end{itemize}


\acknowledgments

We would like thank Paulo Freire and Andrew Matas for useful discussions. AJT is supported by DOE grant DE-FG02-12ER41810. DHW
thanks the Perimeter Institute for hospitality while part of this work was
being completed. AJT and CdR would like to thank the Yukawa Institute for Theoretical Physics at Kyoto University for discussions and hospitality whilst part of this work was being completed, during the YITP-T-12-04 on "Nonlinear massive gravity theory and its observational test".

\begin{center}

\begin{table}
\label{t:DNS}
\hspace{-10pt}
\begin{tabular*}{6.5in}{@{\extracolsep{\fill}} || l | c | c | c | c | c ||}
\hline
 & A & B & C & D & E\\
\hline
Pulsar & 1913+16       & B2127+11 & B1534+12 & J0737--3039 & J1738+0333\\
 & Taylor-Hulse  &          &          & double pulsar & \\
\hline  \hline
$M_1/M_\astrosun$ & 1.386   & 1.358 & 1.345 & 1.338 & 1.46 \\ \hline
$M_2/M_\astrosun$ & 1.442   & 1.354 & 1.333 & 1.249 & 0.181 \\ \hline
$T_P$/days        & 0.323   & 0.335 & 0.420 & 0.102 & 0.355 \\ \hline
$e$               & 0.617   & 0.681 & 0.274 & 0.088 & $3.4\times 10^{-7}$ \\ \hline \hline
$\frac{\d T_P}{\d t}|_{ \pi\ {\rm Monopole}}$  & $4.5 \times 10^{-22}$ & $ 8.3 \times 10^{-22}$ & $1.2 \times 10^{-23}$ & $ 8.1 \times 10^{-25}$
&  $ 2.1 \times 10^{-36}$  \\ \hline
$\frac{\d T_P}{\d t}|_{ \pi\ {\rm Dipole}}$    & $10^{-30}$ & $10^{-32}$ & $10^{-33}$ & $10^{-32}$ & $10^{-31}$\\ \hline
$\frac{\d T_P}{\d t}|_{ \pi\ {\rm Quadrupole}}$& $2.0 \times 10^{-20}$ & $2.2 \times 10^{-20}$ & $1.4 \times 10^{-20}$ & $9.7 \times 10^{-21}$
& $2.4 \times 10^{-21}$\\ \hline
$\frac{\d T_P}{\d t}|_{{\rm GR}}$              & $2.4 \times 10^{-12}$ & $3.8 \times 10^{-12}$ & $1.9 \times 10^{-13}$ & $1.2 \times 10^{-12}$
& $ 2.2\times 10^{-14}$\\ \hline
$\sigma$ &  $5.1\times 10^{-15}$ & $1.3\times 10^{-13}$ &  $2.0\times 10^{-15}$ & $1.7\times 10^{-14}$ & $ 10^{-15}$\\ \hline \hline
Ref. & \cite{Hulse:1974eb,Taylor:1989sw,Weisberg:2004hi}  &  \cite{Jacoby:2006dy} & \cite{Stairs:2002cw,Konacki:2003a} & \cite{Kramer:2006nb} & \cite{Freire:2012mg} \\ \hline
\end{tabular*}
\caption{The predicted
contribution to the orbital period derivative $\dot T_P$
from $\pi$ alone in the monopole, dipole and quadrupole channels (taking $m=1.54\times 10^{-33}$eV) for four known DNS
pulsars (A to D) and one pulsar-white dwarf binary (E) with the GR result. The experimental uncertainty $\sigma$ is given using \cite{Footnote}.}
\end{table}

\end{center}



\begin{thebibliography}{}
\bibitem{Footnote}
We cite $\sigma$ as twice the
formal fitting uncertainty from the TEMPO pulse timing program, where reported.
\end{thebibliography}
\end{document}